\documentclass[epj]{svjour}
\usepackage{graphicx}
\begin{document}
\title{Zeeman energy and anomalous spin splitting in lateral GaAs quantum dots}
\titlerunning{Zeeman energy and anomalous spin splitting}
\author{Manuel Val\'{\i}n-Rodr\'{\i}guez$^1$ \thanks{\email{vdfsmvr4@clust.uib.es}}, 
        Antonio Puente$^1$ \and
        Lloren\c{c} Serra$^{1,2}$}
\authorrunning{M. Val\'{\i}n-Rodr\'{\i}guez {\em et al.}}
\institute{$^1$ Departament de F\'{\i}sica,
Universitat de les Illes Balears, E-07122 Palma de Mallorca, Spain\\*
$^2$ Institut Mediterrani d'Estudis Avan\c{c}ats IMEDEA (CSIC-UIB),
E-07122 Palma de Mallorca, Spain}
\date{March 8, 2004}
\abstract{
The level splittings induced by a horizontal magnetic field in a 
parabolic two-dimensional quantum dot with spin-orbit interaction are 
obtained.
Characteristic features induced by the spin-orbit coupling are  
the appearance of zero-field gaps as well as energy splittings that depend
on the electronic state and the orientation of the magnetic field in the 
quantum-dot plane. It is suggested that these quantum-dot properties could be used to 
determine the Rashba and Dresselhaus spin-orbit intensities}
\maketitle

\section{Introduction} 

Nowadays, spin-related physics has become one of the most active
branches of research in condensed matter. The fundamental physics
involved and the potential applicability in semiconductor device
technology constitute the main reasons encouraging this research.
In particular, there is a special interest in the properties of electronic
spins confined in a quantum dot. The reduced dimensionality of quantum
dots makes them good candidates for systems sustaining long-lived spin 
states \cite{nazarov} and, ultimately, allowing coherent spin 
manipulation \cite{loss}.

A most convenient way to distinguish spin states in a quantum dot
fabricated within a two-dimensional electron gas (2DEG) is  
by means of a magnetic field. When the field is
oriented parallel to the plane of the 2DEG it is expected that 
up and down spin directions will split by the
Zeeman energy, with no additional modification of the electronic states.
This assumption relies on the complete decoupling of the 
orbital motion and the parallel magnetic field. Since orbital motion 
is restricted to the 2DEG's plane, the vertical Lorentz force associated 
with a horizontal field becomes irrelevant. However, a 
qualitative difference appears when the electron feels an important 
spin-orbit (SO) interaction during its orbital motion.
In this case the SO mechanism is effectively coupling the electronic states
with the horizontal magnetic field, beyond a pure Zeeman splitting, even 
in two-dimensional dots.
 
Recently, several groups have measured the
level splittings of lateral GaAs quantum dots in a parallel magnetic field
\cite{hanson,potok}, even for a single electron occupation of the dot. 
The measured splittings are usually interpreted 
in terms of an effective $g$-factor in a simple Zeeman formula.
It is the aim of this work to analyze the theoretical prediction of
level splittings for a quantum dot in a parallel magnetic field
when SO interaction is present, 
emphasizing the deviations from the Zeeman scenario.
It is also our purpose to discuss the possible SO signatures in 
the recent experimental data for single-electron dots \cite{hanson,potok}.

\section{Quantum dot model} 

We consider a 2D 
representation of the effective mass Hamiltonian for the GaAs 
conduction-band electrons ($m^*=0.067m_e$)
\begin{equation}
{\cal H}_{xy}=\frac{p_x^2+p_y^2}{2m^*}+
\frac{1}{2}m^*\omega_0^2(x^2+y^2)\; .
\end{equation}
The above hypothesis implies 
that the direction perpendicular to the 2DEG is strongly quantized in 
comparison with the in-plane degrees of freedom.
The potential responsible for the confinement
in the plane is taken as an isotropic parabola of 
constant $\omega_0$.
We also include the Zeeman energy
\begin{equation}
{\cal H}_Z=\frac{1}{2}g^*\mu_B \vec{B}\cdot\vec{\sigma}\; ,
\end{equation}
where $\vec{B}\equiv(B_x,B_y)$ and $\vec{\sigma}\equiv(\sigma_x,\sigma_y)$ 
are the magnetic field
and the Pauli-matrix vectors, respectively; $\mu_B$ is the Bohr magneton
and $g^*=-0.44$ is the bulk GaAs $g$-factor. 

The SO interaction is taken into account by adding the linear
Dresselhaus \cite{dress} and Bychkov-Rashba \cite{rash} terms
for conduction band electrons in a [001] 2DEG \cite{knap},
\begin{eqnarray}
{\cal H}_D &=& \frac{\lambda_D}{\hbar}(p_x\sigma_x-p_y\sigma_y)\; , \\
{\cal H}_R &=& \frac{\lambda_R}{\hbar}(p_y\sigma_x-p_x\sigma_y)\; .
\end{eqnarray}
These contributions originate, as a relativistic effect, in the 
electric fields present in the heterostructure. In their intrinsic 
reference frame, the moving electrons feel 
these electric fields as effective magnetic fields that interact 
with their spin.
The Dresselhaus term is due to  
the bulk inversion asymmetry of the GaAs crystal and its intensity 
depends on the expectation value of the vertical momentum 
and a material-dependent constant ($\gamma=27.5$ eV\AA$^3$ for GaAs) as
\begin{equation}
\lambda_D = \gamma \langle k_z^2\rangle \approx \gamma \left(
\frac{\pi}{z_0}\right)^2\; ,
\end{equation} 
where $z_0$ is the vertical width of the 2DEG. 
On the other hand,
the Rashba contribution stems from the asymmetry of the heterostructure 
profile (the built-in electric field) and it is also sensitive
to the electric fields induced by external gates \cite{grun}.
Below we shall treat $\lambda_D$ and $\lambda_R$ as varying parameters
and study the results for different values. From Shubnikov-de Haas
as well as magnetotransport measurements in 2DEG's values for these
parameters ranging
from $\approx 5$ meV{\AA} to $\approx 50$ meV{\AA} 
have been inferred in GaAs\cite{ramvall,miller}.

Adding all the above contributions, the full Hamiltonian ${\cal H}$ can be considered 
as composed of two parts.
The first one (${\cal H}_0$), containing the kinetic energy, the confining
potential and the SO interaction, remains invariant by the 
time-reversal symmetry.
Conversely, the second part of the Hamiltonian, corresponding to the Zeeman
term ${\cal H}_Z$, breaks this symmetry. More specifically, we have
\begin{eqnarray}
\cal{H} &=& {\cal H}_0 + {\cal H}_Z\; ,\\
{\cal H}_0 &=& {\cal H}_{xy} + {\cal H}_D + {\cal H}_R\; . 
\end{eqnarray}

\section{Analytical approximations}

The analytical diagonalization of the full Hamiltonian is not available, 
although several analytical results can be obtained 
when different pieces of the Hamiltonian are neglected or
treated within perturbation theory. The perturbative 
analysis when Zeeman and SO terms are of the same order of magnitude 
is rather involved and this case will be considered only numerically.
In this section we shall consider the two limiting cases of 
zero Zeeman energy with a relatively weak SO interaction, and 
of a Zeeman energy much larger than SO effects.  

Of course, another limit with analytical solution appears when the 
Zeeman energy is included and the SO interaction is switched off. 
Under these assumptions the Hamiltonian is trivial and its 
level structure is that of a
two-dimensional oscillator 
\begin{eqnarray}
\label{eq8a}
\varepsilon_{n{\ell}s} &=& (2n+\left| \ell \right| + 1)\hbar\omega_0
+\frac{1}{2}g^*\mu_BB s
\; ,
\end{eqnarray}
whose states are spin-split by the Zeeman 
energy gap $\Delta_s=|g^*|\mu_BB$, independently of the orbital state 
considered. In Eq.\ (\ref{eq8a}), $n=0,1,\dots$ and $\ell=0,\pm1,\dots$ 
are the principal and $L_z$  quantum numbers, respectively, 
while $s=\pm 1$ is the spin label.

\subsection{Limit of weak SO in zero field}

Assuming ${\cal H}_Z=0$ a perturbative calculation for 
${\cal H}_{R,D}<<\hbar\omega_0$ yields corrections 
of second order in the $\lambda$'s and, due
to the oscillator and spin degeneracies, requires degenerate 
perturbation theory. The modified energy levels read
\begin{eqnarray}
\label{eq8}
\varepsilon_{n{\ell}s} &=& (2n+\left| \ell \right| + 1)\hbar\omega_0
\nonumber\\
&-& \frac{m^*}{\hbar^2}
(\lambda_D^2+\lambda_R^2) +
\frac{m^*}{\hbar^2}
(\lambda_D^2-\lambda_R^2){\ell}s\; .
\end{eqnarray}

The level structure given by Eq.\ (\ref{eq8}) 
corresponds to that of a two-dimensional oscillator with a
constant energy shift and a
fine structure depending on $\ell$ and $s$.
Both shift and splitting are proportional to the 
SO intensities.
It is worth stressing that even in the present case of negligible
Zeeman energy the SO interaction yields a spin-splitting of the 
energy levels given by 
$\Delta_s=2\frac{m^*}{\hbar^2}\left|\, \ell (\lambda_D^2-\lambda_R^2)\,\right|$. 
This zero-field splitting is consistent with the Kramers degeneracy 
present in a half-integer-spin system with time reversal symmetry. 
Indeed, each eigenstate
characterized by the values of ($n,\ell,s$) has a 
conjugate ($n,-\ell,-s$) at the same energy.

An alternative method to derive Eq.\ (\ref{eq8}) was suggested in 
Ref.\ \cite{alei} and uses a unitary transformation of ${\cal H}_0$
leading to a new Hamiltonian that is diagonal in spin space up to second 
order in the SO intensities. The
remaining non-diagonal terms of $O(\lambda^3)$ are small compared with the
diagonalized ones for typical GaAs $\lambda$'s, and so 
they can be dropped without significant error.
Using this alternative transformation we analyzed quantum dot properties like 
the far-infrared absorption in Ref.\ \cite{valinfir} and the spin 
precession in Ref.\ \cite{valinprec}.

\subsection{Limit of weak SO in large field}

We assume now that there is a large horizontal field, such that 
${\cal H}_{R,D}<<{\cal H}_Z$. The unperturbed 
energy levels are those of Eq.\ (\ref{eq8a}), where up and down 
spin is defined in the direction of the magnetic field
$\vec{B}\equiv B(\cos\theta,\sin\theta)$. As before, 
a second-order calculation within degenerate perturbation theory 
yields
\begin{eqnarray}
\label{eq2o}
\varepsilon_{n{\ell}s} = \varepsilon_{n{\ell}s}^{(0)} 
&-& \frac{m^*}{2\hbar^2}\left\{\rule{0mm}{5mm}\right. G  
+ F \frac{1}{1-z^2}\nonumber\\
&\times& \left[\rule{0mm}{4mm}
1+s(2n+\ell+|\ell|+1)z
\right]
\left.\rule{0mm}{5mm}\right\}\; ,
\end{eqnarray}
where $\varepsilon_{n{\ell}s}^{(0)}$ is given by Eq.\ (\ref{eq8a}),
and we have defined $z=g^*\mu_B B/\hbar\omega_0$, the ratio of Zeeman 
energy to external confinement frequency, as well as the two auxiliary 
quantities
\begin{eqnarray}
G &=& \lambda_R^2 + \lambda_D^2 - 2\lambda_R\lambda_D\sin{(2\theta)}\; ,\\
F &=& \lambda_R^2 + \lambda_D^2 + 2\lambda_R\lambda_D\sin{(2\theta)}\; .
\end{eqnarray}

\begin{figure}[t]
\centerline{\includegraphics[width=3.25in,clip=]{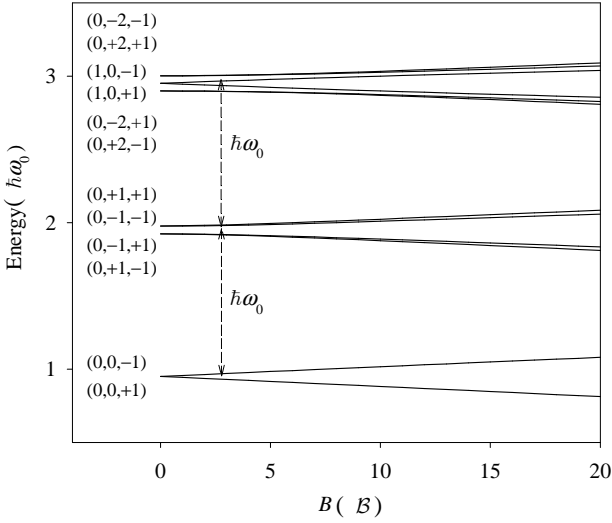}}
\caption{Magnetic-field evolution of the first three oscillator shells of 
a parabolic quantum dot with SO interaction. 
The magnetic field is along the direction $\theta=45$ deg and the SO 
intensities are
$\lambda_D=0.2\,\hbar\omega_0\ell_0$ and $\lambda_R=0.1\,\hbar\omega_0\ell_0$. 
The labels 
indicate the values of $(n,\ell,s)$ for each $B=0$ level (see text).}
\end{figure}

In Eq.\ (\ref{eq2o}) the degeneracy of each  
major shell, with a given $N=2n+|\ell|$, is broken into multiplets
of states, as will be further clarified below when discussing the numerical
results. It can be shown that the energies for $n=0$, negative $\ell$ 
and $s=+1$ smoothly converge to the exact 
values for large enough magnetic fields. For other values of these 
labels there are discontinuities in the range of validity of Eq.\ (\ref{eq2o}) 
since an additional restriction is that the field, besides of being large
enough, should not be close to 
a crossing point with an integer $z$. Note that 
Eq.\ (\ref{eq2o}) may actually diverge for $z=1$. Another 
relevant feature of Eq.\ (\ref{eq2o}) is that it takes
into account the angular anisotropy through the $\sin{(2\theta)}$ 
contribution to $F$ and $G$.
   
It should be emphasized that although the energy levels from the above 
perturbative calculations, Eqs.\ (\ref{eq8}) and (\ref{eq2o}), can still 
be classified using the uperturbed labels $(n,\ell,s)$, the 
corresponding states are no longer eigenstates of orbital and spin 
angular momenta. These results generalize those of Ref.\ \cite{Gov02}
by including the two sources of SO coupling as well as the arbitrary
angular orientation of the horizontal magnetic field.

\section{Numerical solutions}

As stated above, when both SO and Zeeman terms are included 
on the same footing a numerical resolution is
required. We have used two 
alternative methods: a) by spatial discretization in a 
a uniform two-dimensional grid, and b) using an oscillator basis.
In both methods we do not impose any symmetry restriction in either 
real or spin spaces and we have checked that the above
analytical limits are recovered with high accuracy.

\begin{figure}[t]
\centerline{\includegraphics[width=2.8in,clip=]{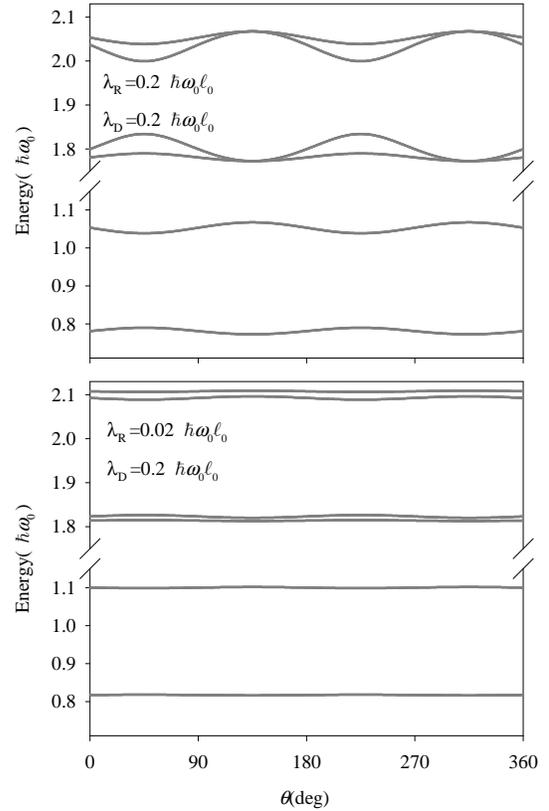}}
\caption{Angular anisotropy of the first and second shells for
a magnetic field of modulus $B=20 {\cal B}$ and the 
SO parameters displayed in each panel. The angle $\theta$ is the 
polar angle of the magnetic field vector.}
\end{figure}

A natural unit system for the calculations is given by the parabola
energy $\hbar\omega_0$ and the oscillator length 
$\ell_0^2=\hbar/(m^*\omega_0)$.
The associated unit of magnetic field is 
${\cal B}=\omega_0 m^*c/e$ while the SO intensities
are given in units of $\hbar\omega_0\ell_0$. Assuming, for instance, 
a GaAs parabolic dot with $\hbar\omega_0= 1$ meV and $\lambda_D=50$ meV{\AA}
we would then have $\ell_0\approx 338$ \AA, 
${\cal B}\approx 0.58$ T and, thus, 
$\lambda_D\approx 0.15\,\hbar\omega_0\ell_0$.

Figure 1 shows a typical dependence of a quantum dot's level structure 
with an
applied horizontal magnetic field for the three lowest
harmonic oscillator shells. The lowest subband corresponds to states
having ($n=0,\ell=0,s=\pm1$) and it can
be seen that these two states split linearly with the applied 
magnetic 
field, in a similar way to a usual Zeeman splitting. 

The second subband is composed
of four states characterized by ($n=0,\ell=\pm1,s=\pm1$). These states 
show a zero-field spin-splitting and at large fields they group
into two branches, each of them composed of two almost parallel 
close lines. 
In the intermediate-field regime the evolution of 
these two branches is nonlinear with $B$.
Note that the labels are given in the left part of the diagram according
to the ordering of Eq.\ (\ref{eq8}), which is appropriate to the $B=0$ limit.
At high $B$'s the ordering, given by Eq.\ (\ref{eq2o}), changes 
since $s=+1$ ($-1$) corresponds to the lower (upper) branch.

The third subband includes both 
linear and non-linear splittings since this subband is composed of 
two states ($n=1,\ell=0,s=\pm1$) that split linearly 
with $B$, and four states 
($n=0,\ell=\pm2,s=\pm1$) with a zero-field spin-splitting and a 
non-linear evolution. In general, at low fields states with $\ell=0$ always
split linearly with $B$ (thus in a Zeeman-like way) while states having 
$\ell\neq 0$ do not split to first order in the field. 

It can be easily checked that the 
the analytical approximations discussed in Sec.\ III
reproduce the zero-field 
splittings and the fine-structured branches at large $B$ in Fig.\ 1.
The representation in this figure is a generalization to the parallel 
case
of the well-known Fock-Darwin diagrams for quantum dots in perpendicular 
fields.
When calculating this type of diagrams it is necessary to
specify the orientation of the parallel magnetic field
in the 2DEG plane since, in general, the anisotropy of the SO terms
can reflect in an angular dependence of the level structure.
However, there is one case
where the level structure is fully isotropic and it is when only one 
source of SO coupling is present (Dresselhaus or Bychkov-Rashba). 
A general proof of this statement can be found in the following.
Note, however, that the perturbative result Eq.\ (\ref{eq2o})
already predicts that the level structure is $\theta$-independent 
when one of the two $\lambda$'s vanishes. 

If only one source of SO interaction is present in the Hamiltonian, 
its part preserving time reversal symmetry (${\cal H}_0$) fulfills
another continuous symmetry ${\cal S}$ with generator ${\cal G}$; 
i.e., ${\cal S}=e^{-i\theta{\cal G}/\hbar}$. 
For the Bychkov-Rashba and Dresselhaus cases it is ${\cal G}=L_z+S_z$
and ${\cal G}=L_z-S_z$, respectively. This result follows inmediately
from the fact that both ${\cal H}_R$ and ${\cal H}_D$ commute with their
corresponding symmetry generators. Taking into account that 
the energy levels of ${\cal H}$ and ${\cal S}^+{\cal H}{\cal S}$
are identical and that the latter corresponds in fact to a rotation  
of the magnetic field around a vertical axis and angle $\theta$ 
it follows that the level structure does not depend on the parallel
field orientation.
The transformed Hamiltonian indeed corresponds to a rotated field since
\begin{eqnarray}
{\cal S}^+{\cal H}{\cal S}&=&
{\cal H}_0 + \frac{1}{2}g^* \mu_B\, \vec{B}\cdot 
\left(e^{i\theta S_z}\, {\vec\sigma}\, e^{-i\theta S_z}\right)\nonumber\\
&=& {\cal H}_0 + \frac{1}{2}g^* \mu_B\, \vec{B}_r\cdot\vec{\sigma}\; ,
\end{eqnarray}
where $\vec{B}_r=(B_x\cos\theta+B_y\sin\theta, B_y\cos\theta-B_x\sin\theta)$
is the rotated field.
This argument breaks down
when both SO interaction sources are present since, then, the continuous
symmetry ${\cal S}$ is lost and as a consequence, the level structure
is anisotropic. 

In agreement with the above discussion, the numerical calculations
show that the angular anisotropy of the energy levels is maximal 
when $\lambda_R\sim\lambda_D$ and negligible when one SO intensity 
is much smaller than the other. This is clearly seen in 
Fig.\ 2, which displays the evolution with the magnetic
field orientation of the levels corresponding to the lowest and the 
first-excited subbands of a dot in a parallel field 
$B=20 {\cal B}$. 
It is worth mentioning that not only the relative 
weight $\lambda_R/\lambda_D\sim 1$ yields an important anisotropy 
of the level structure.
The abolute values are also relevant since, obviously, when both 
intensities are exceedingly smaller than $\hbar\omega_0\ell_0$  
SO effects become negligible and the level structure is
isotropic irrespectively of the relative weight.
Another interesting feature is that the level anisotropy is sensitive
to the relative sign of $\lambda_R$ and $\lambda_D$. This is at variance
with most physical properties, like the $\theta$-averaged 
values of the energy levels, that are depending only on the absolute values 
$|\lambda_R|$ and $|\lambda_D|$. Focussing, for instance, on the first 
shell gap, we find from the numerical calculations 
that if $\lambda_R\lambda_D>0$ the gap minima 
are at $\theta=\pi/4+m{\pi}$ (with $m$ an integer) 
while if $\lambda_R\lambda_D<0$ the gap
minima are shifted to  
$\theta=3\pi/4+m{\pi}$. In all cases, the anisotropy is characterized
by a periodicity in $2\theta$, in agreement with the analytical formula
derived above.

\begin{figure}[b]
\centerline{\includegraphics[width=2.5in,clip=]{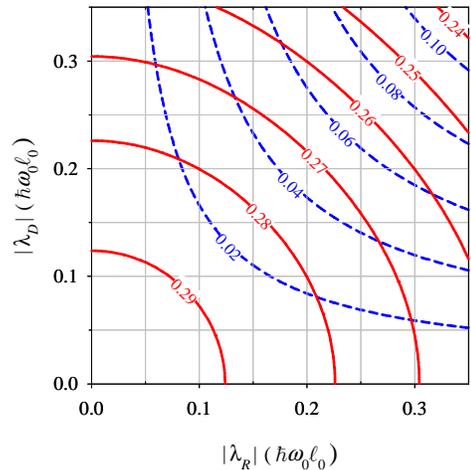}}
\caption{Contour plots with the dependence on the spin-orbit intensities of the 
first shell-splitting at $B=20 {\cal B}$. Solid lines
show the $\theta$-averaged 
splitting $\bar\Delta_s$ in $\hbar\omega_0$ units (the 
Zeeman energy in these units is $E_Z=0.2948$).
Dashed lines correspond to the amplitude of the $\theta$ oscillation,
$\left[\max(\Delta_s(\theta))-\min(\Delta_s(\theta))\right]$.}
\end{figure}
\begin{figure}[t]
\centerline{\includegraphics[width=2.5in,clip=]{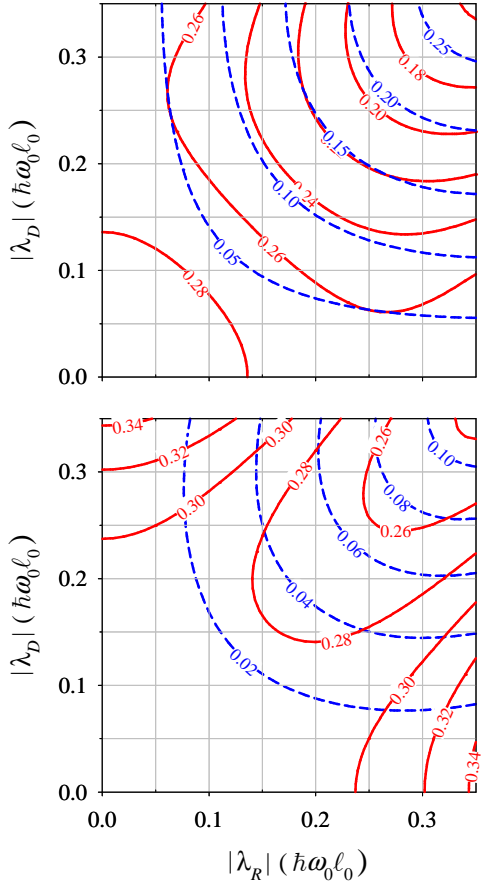}}
\caption{Same as Fig.\ 3 for the second-shell splittings.
Lower panel corresponds to the energy gap between higher and lower states,
while upper panel shows the gap between the two intermediate ones.}
\end{figure}
 
The detailed variation of the first-shell splitting
with the SO intensities when $B=20 {\cal B}$ is summarized in Fig.\ 3. 
We note that for the chosen range of $\lambda$'s the deviation 
of the $\theta$-averaged splitting $\bar\Delta_s$ from the 
Zeeman value ($0.2948\hbar\omega_0$ at this $B$) ranges from 0 to  
$\approx 20$\%. 
The angular anisotropy (dashed lines) is measured by the difference between maximum
and minimum energy gap, i.e.,  
$\left[\max(\Delta_s(\theta))-\min(\Delta_s(\theta))\right]$
and, in agreement with the above discussion, reaches maximum values 
along the diagonal line $|\lambda_R|=|\lambda_D|$, ranging from
zero to $\approx 50$\% of $\bar\Delta_s$.
Figure 4 shows similar results for the  
splitting of the second-shell. Notice that in this more complex case 
we define two splittings:  one gives the energy
difference between the higher and lower states of the second shell
while the other is associated with the two intermediate ones.

The numerical results in Fig.\ 3 qualitatively agree with the prediction
from Eq.\ (\ref{eq2o}) of an averaged splitting and anisotropy  
proportional to $(\lambda_R^2+\lambda_D^2)$ and $\lambda_R\lambda_D$, 
respectively. Nevertheless, the difference between the 
actual values from the second-order
and numerical calculations increases with the $\lambda$'s and it 
leads to sizeable errors for values above $\sim 0.2$.
For the second-shell results (Fig.\ 4) the errors from the 
perturbative calculation are more conspicuous since they lead
to qualitatively different contour lines for $\lambda$'s above $\sim 0.15$.

The above results suggest that a high-precission measurement of the quantum 
dot level structure could determine the SO intensities.
One should know the specific values of the underlying 2DEG (like $g^*$ and $m^*$)
as well as the value of $\omega_0$. The zero-field splitting
of the second shell would fix $|\lambda_D^2-\lambda_R^2|$.
This condition along with the results of the 
first-shell splitting at large $B$ ($\bar\Delta_s$ in Fig.\ 3) 
would yield the lower and greater SO intensities in absolute value,
i.e., $|\lambda_<|$ and $|\lambda_>|$.
The angular anisotropy of the splittings could fix,
as mentioned above, the relative sign.
To discern which source (Bychkov-Rashba or Dresselhaus) is the greater or lower
SO intensity seems a somewhat complicate task. In principle, however, it could
be accomplished by looking at the second shell splittings since when 
$|\lambda_D|>|\lambda_R|$
the upper (lower) states of the second shell have parallel (antiparallel) spin and 
orbital angular momenta at zero field. 

The role of the orbital level spacing is quite important for understanding the
$B$-evolution of the spin splittings when SO interaction is present.
To see this, we display in Fig.\ 5 the lowest subband's splitting
when the SO intensitites are kept constant and the confining frequencies
are varied.
Small dots, characterized by a wide gap between
orbital subbands, show no effect of the SO interaction in this
splitting, but, as the orbital and spin energy scales become comparable,
due to the proximity of the orbital subbands, SO interaction induces
a level repulsion that reflects in a compression of the levels reducing
the spin splitting. For the particular $\lambda$'s and magnetic field
orientation in Fig.\ 5, the 
dots characterized by an orbital level spacing lower than $\approx 1$ meV 
there is a sizeable reduction of the splitting with respect to the Zeeman
value and thus their effective $g$-factor will be lower
than that of the 2DEG.

\begin{figure}[t]
\centerline{\includegraphics[width=2.5in,clip=]{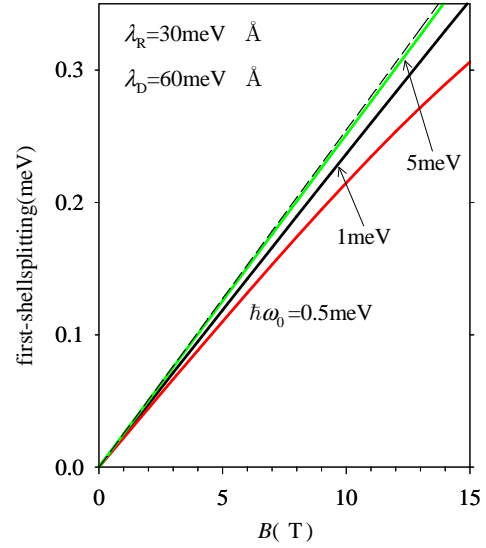}}
\caption{Evolution with $B$ of the lower shell splitting
for  fixed SO intensities and different values of the confinement
$\hbar\omega_0$. The magnetic field orientation in the 2DEG's plane
is given by a polar angle of $\theta=45$ deg. The dashed line shows the 
Zeeman energy
$|g^*|\mu_B B$.}
\end{figure}

The B-evolution corresponding to the second shell is more complex
since, as already discussed above, this subband involves four different states.
Figure 6 shows the two energy splittings corresponding
to the gap between intermediate states and from the lowest to the highest one,
respectively, of the second shell.
In all cases, the evolution of the splittings is clearly non-linear showing a positive
curvature at low $B$'s and the above-mentioned
finite value at zero field.
It can be seen that as the orbital level spacing is reduced
the values of the splittings at large $B$ are also reduced
due to the level repulsion induced by SO interaction,
similarly to the case of the first subband. 
A very interesting feature is that for low enough $\hbar\omega_0$ the 
SO coupling induces a large separation of the two $B$-dependent 
gaps. A difference that does not appear in the case of small dots (high
$\omega_0$'s) and, thus, it reveals the SO-induced fine structure of the 
dot level spectrum.

\begin{figure}[t]
\centerline{\includegraphics[width=3.25in,clip=]{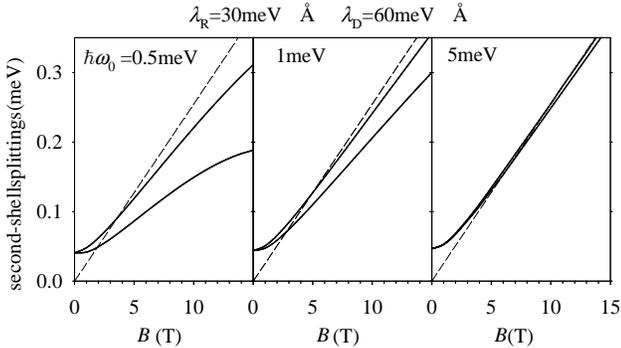}}
\caption{Same as Fig.\ 5 for the energy splittings of the second shell.
One splitting gives the energy gap between
the highest and lowest states of this shell, while the other corresponds
to the difference between the two intermediate states.}
\end{figure}

\section{Discussion on experimental evidences and conclusions}

In a recent work Potok {\em et al} \cite{potok} have measured 
the level structure and the corresponding spin splittings of the
lowest and first-excited subbands for a GaAs lateral
dot in a horizontal magnetic field
(Fig.\ 2c of Ref.\ \cite{potok}).
The data show a small deviation of the first-shell splitting
at large $B_\parallel$ from the pure Zeeman result, with a fitted value of 
$|g|\sim 0.37$ (to be compared with $0.44$ for the bulk). 
A single value of the second-shell splitting at large $B_\parallel$
is reported, 
since the resolution
does not seem enough to discriminate finer structures.
This second-shell energy gap is found to be somewhat 
lower than that of the first-shell. Unfortunately, 
the angular dependence is not discussed and the
zero-field splitting of the second shell is not resolved, although 
the data suggest the existence of unresolved structures.
The value of the 
confining frequency inferred from the level 
spacing is $\hbar\omega_0\approx 0.8$ meV.
As shown in Figs.\ 5 and 6, SO-induced effects for $\lambda$'s
in the range 30 to 60 meV{\AA} can reproduce the observations, but only 
for a selected magnetic field orientation. The corresponding angular 
anisotropies inferred from Figs.\ 3 and 4 are important
and would reduce/enhance the difference from the pure Zeeman value 
when varying the polar angle $\theta$.  

Hanson {\em et al} \cite{hanson} have also measured the spin-splitting 
of the first shell of a GaAs parabolic dot, this case evidencing a clear
nonlinear $B_\parallel$-dependence
(see Fig.\ 1e of Ref.\ \cite{hanson}).
The deviation from linear behavior is  
qualitatively similar to the $\hbar\omega_0=0.5$ meV results 
of Fig.\ 5 although the experimental value of $\hbar\omega_0$ at low $B_\parallel$'s
is close to 1 meV. Therefore, to ascribe the observation to SO effects 
one should assume that 
the value of the confinement $\hbar\omega_0$ decreases as 
$B_\parallel$ increases\cite{valinpd}.

In summary, we have analyzed the energy level structure and splittings 
that are predicted
within the effective-mass model of a 2D parabolic quantum dot with  
Bychkov-Rashba and Dresselhaus SO interactions. The deviations from the
simple Zeeman-splitting scenario have been stressed. Characteristic 
features of the SO interactions are the appearance of zero-field splittings,
effective $g$-factors smaller than the bare one and sizeable
anisotropies when both SO sources are of the same order.
It has been suggested that these features could be used to
determine the SO intensities from quantum dot measurements.

Spin-orbit intensities in the range of $\approx 50$ meV{\AA} yield effects
that are close to recent observations in GaAs quantum dots.
However, a conclusive evidence
allowing to validate or discard the relevance of the SO interaction for 
these measurements is lacking. In this respect, the clarification of  
the experimental zero-field splitting of the second shell as well as of
the $\theta$-dependence of the splittings would be very important.

This work was supported by Grant No.\ BFM2002-03241 
from DGI (Spain).

\end{document}